\documentclass[a4paper,11pt]{article}
\pdfoutput=1 % if your are submitting a pdflatex (i.e. if you have
\usepackage[T1]{fontenc} % if needed

\makeatletter
\gdef\@fpheader{ }
\gdef\@journal{ }
\makeatother
\RequirePackage{amsmath}
\RequirePackage{amssymb}
\RequirePackage{epsfig}
\RequirePackage{graphicx}
\RequirePackage[numbers,sort&compress]{natbib}
\RequirePackage{color}
\RequirePackage[colorlinks=true
,urlcolor=blue
,anchorcolor=blue
,citecolor=blue
,filecolor=blue
,linkcolor=blue
,menucolor=blue
,pagecolor=blue
,linktocpage=true
,pdfproducer=medialab
,pdfa=true
]{hyperref}

\newif\ifnotoc\notocfalse
\newif\ifemailadd\emailaddfalse
\newif\iftoccontinuous\toccontinuousfalse
\makeatletter
\def\@subheader{\@empty}
\def\@keywords{\@empty}
\def\@abstract{\@empty}
\def\@xtum{\@empty}
\def\@dedicated{\@empty}
\def\@arxivnumber{\@empty}
\def\@collaboration{\@empty}
\def\@collaborationImg{\@empty}
\def\@proceeding{\@empty}
\def\@preprint{\@empty}

\newcommand{\subheader}[1]{\gdef\@subheader{#1}}
\newcommand{\keywords}[1]{\if!\@keywords!\gdef\@keywords{#1}\else%
\PackageWarningNoLine{\jname}{Keywords already defined.\MessageBreak Ignoring last definition.}\fi}
\renewcommand{\abstract}[1]{\gdef\@abstract{#1}}
\newcommand{\dedicated}[1]{\gdef\@dedicated{#1}}
\newcommand{\arxivnumber}[1]{\gdef\@arxivnumber{#1}}
\newcommand{\proceeding}[1]{\gdef\@proceeding{#1}}
\newcommand{\xtumfont}[1]{\textsc{#1}}
\newcommand{\correctionref}[3]{\gdef\@xtum{\xtumfont{#1} \href{#2}{#3}}}
\newcommand\jname{JHEP}
\newcommand\acknowledgments{\section*{Acknowledgments}}

\newcommand\preprint[1]{\gdef\@preprint{\hfill #1}}

\newenvironment{proof}[1][Proof]{\noindent\textbf{#1.} }{\ \rule{0.5em}{0.5em}}

\makeatother

\newcommand\note[2][]{%
\if!#1!%
\stepcounter{footnote}\footnotetext{#2}%
\else%
{\renewcommand\thefootnote{#1}%
\footnotetext{#2}}%
\fi}

\makeatletter
\newtoks\auth@toks
\renewcommand{\author}[2][]{%
  \if!#1!%
    \auth@toks=\expandafter{\the\auth@toks#2\ }%
  \else
    \auth@toks=\expandafter{\the\auth@toks#2$^{#1}$\ }%
  \fi
}
\makeatother
\makeatletter
\newtoks\affil@toks\newif\ifaffil\affilfalse
\newcommand{\affiliation}[2][]{%
\affiltrue
  \if!#1!%
    \affil@toks=\expandafter{\the\affil@toks{\item[]#2}}%
  \else
    \affil@toks=\expandafter{\the\affil@toks{\item[$^{#1}$]#2}}%
  \fi
}
\makeatother
\makeatletter
\newtoks\email@toks\newcounter{email@counter}%
\newcommand{\emailAdd}[1]{%
\emailaddtrue%
\ifnum\theemail@counter>0\email@toks=\expandafter{\the\email@toks, \@email{#1}}%
\else\email@toks=\expandafter{\the\email@toks\@email{#1}}%
\fi\stepcounter{email@counter}}
\newcommand{\@email}[1]{\href{mailto:#1}{\tt #1}}
\makeatother

\makeatletter
\newcommand*\collaboration[1]{\gdef\@collaboration{#1}}
\newcommand*\collaborationImg[2][]{\gdef\@collaborationImg{#2}}
\makeatletter
\newcommand\afterLogoSpace{\smallskip}
\newcommand\afterSubheaderSpace{\vskip3pt plus 2pt minus 1pt}
\newcommand\afterProceedingsSpace{\vskip21pt plus0.4fil minus15pt}
\newcommand\afterTitleSpace{\vskip23pt plus0.06fil minus13pt}
\newcommand\afterRuleSpace{\vskip23pt plus0.06fil minus13pt}
\newcommand\afterCollaborationSpace{\vskip3pt plus 2pt minus 1pt}
\newcommand\afterCollaborationImgSpace{\vskip3pt plus 2pt minus 1pt}
\newcommand\afterAuthorSpace{\vskip5pt plus4pt minus4pt}
\newcommand\afterAffiliationSpace{\vskip3pt plus3pt}
\newcommand\afterEmailSpace{\vskip16pt plus9pt minus10pt\filbreak}
\newcommand\afterXtumSpace{\par\bigskip}
\newcommand\afterAbstractSpace{\vskip16pt plus9pt minus13pt}
\newcommand\afterKeywordsSpace{\vskip16pt plus9pt minus13pt}
\newcommand\afterArxivSpace{\vskip3pt plus0.01fil minus10pt}
\newcommand\afterDedicatedSpace{\vskip0pt plus0.01fil}
\newcommand\afterTocSpace{\bigskip\medskip}
\newcommand\afterTocRuleSpace{\bigskip\bigskip}
\newlength{\affiliationsSep}
\newcommand\beforetochook{\pagestyle{myplain}\pagenumbering{roman}}

\DeclareFixedFont\trfont{OT1}{phv}{b}{sc}{11}

\renewcommand\maketitle{
\pagestyle{empty}
\thispagestyle{titlepage}
\setcounter{page}{0}
\noindent{\small\scshape\@fpheader}\@preprint\par

\afterLogoSpace
\if!\@subheader!\else\noindent{\trfont{\@subheader}}\fi
\afterSubheaderSpace
\if!\@proceeding!\else\noindent{\sc\@proceeding}\fi
\afterProceedingsSpace
{\LARGE\flushleft\sffamily\bfseries\@title\par}
\afterTitleSpace
\hrule height 1.5\p@%
\afterRuleSpace
\if!\@collaboration!\else
{\Large\bfseries\sffamily\raggedright\@collaboration}\par
\afterCollaborationSpace
\fi
\if!\@collaborationImg!\else
{\normalsize\bfseries\sffamily\raggedright\@collaborationImg}\par
\afterCollaborationImgSpace
\fi
{\bfseries\raggedright\sffamily\the\auth@toks\par}
\afterAuthorSpace
\ifaffil\begin{list}{}{%
\setlength{\leftmargin}{0.28cm}%
\setlength{\labelsep}{0pt}%
\setlength{\itemsep}{\affiliationsSep}%
\setlength{\topsep}{-\parskip}}
\itshape\small%
\the\affil@toks
\end{list}\fi
\afterAffiliationSpace
\ifemailadd %% if emailadd is true
\noindent\hspace{0.28cm}\begin{minipage}[l]{.9\textwidth}
\begin{flushleft}
\textit{E-mail:} \the\email@toks
\end{flushleft}
\end{minipage}
\else %% if emailaddfalse do nothing
\PackageWarningNoLine{\jname}{E-mails are missing.\MessageBreak Plese use \protect\emailAdd\space macro to provide e-mails.}
\fi
\afterEmailSpace
\if!\@xtum!\else\noindent{\@xtum}\afterXtumSpace\fi
\if!\@abstract!\else\noindent{\renewcommand\baselinestretch{.9}\textsc{Abstract:}}\ \@abstract\afterAbstractSpace\fi
\if!\@keywords!\else\noindent{\textsc{Keywords:}} \@keywords\afterKeywordsSpace\fi
\if!\@arxivnumber!\else\noindent{\textsc{ArXiv ePrint:}} \href{http://arxiv.org/abs/\@arxivnumber}{\@arxivnumber}\afterArxivSpace\fi
\if!\@dedicated!\else\vbox{\small\it\raggedleft\@dedicated}\afterDedicatedSpace\fi
\ifnotoc\else
\iftoccontinuous\else\newpage\fi
\beforetochook\hrule
\tableofcontents
\afterTocSpace
\hrule
\afterTocRuleSpace
\fi
\setcounter{footnote}{0}
\pagestyle{myplain}\pagenumbering{arabic}
} % close the \renewcommand\maketitle{

\renewcommand{\baselinestretch}{1.1}\normalsize
\divide\@tempdima\baselineskip
\@tempcnta=\@tempdima
\skip\@mpfootins = \skip\footins
\renewcommand{\@dotsep}{10000}

\newcommand\ps@myplain{
\pagenumbering{arabic}
\renewcommand\@oddfoot{\hfill-- \thepage\ --\hfill}
\renewcommand\@oddhead{}}
\let\ps@plain=\ps@myplain

\newcommand\ps@titlepage{\renewcommand\@oddfoot{}\renewcommand\@oddhead{}}

\numberwithin{equation}{section}

\renewcommand\section{\@startsection{section}{1}{\z@}%
                                   {-3.5ex \@plus -1.3ex \@minus -.7ex}%
                                   {2.3ex \@plus.4ex \@minus .4ex}%
                                   {\normalfont\large\bfseries}}
\renewcommand\subsection{\@startsection{subsection}{2}{\z@}%
                                   {-2.3ex\@plus -1ex \@minus -.5ex}%
                                   {1.2ex \@plus .3ex \@minus .3ex}%
                                   {\normalfont\normalsize\bfseries}}
\renewcommand\subsubsection{\@startsection{subsubsection}{3}{\z@}%
                                   {-2.3ex\@plus -1ex \@minus -.5ex}%
                                   {1ex \@plus .2ex \@minus .2ex}%
                                   {\normalfont\normalsize\bfseries}}
\renewcommand\paragraph{\@startsection{paragraph}{4}{\z@}%
                                   {1.75ex \@plus1ex \@minus.2ex}%
                                   {-1em}%
                                   {\normalfont\normalsize\bfseries}}
\renewcommand\subparagraph{\@startsection{subparagraph}{5}{\parindent}%
                                   {1.75ex \@plus1ex \@minus .2ex}%
                                   {-1em}%
                                   {\normalfont\normalsize\bfseries}}

\def\fnum@figure{\textbf{\figurename\nobreakspace\thefigure}}
\def\fnum@table{\textbf{\tablename\nobreakspace\thetable}}

\long\def\@makecaption#1#2{%
  \vskip\abovecaptionskip
  \sbox\@tempboxa{\small #1. #2}%
  \ifdim \wd\@tempboxa >\hsize
    \small #1. #2\par
  \else
    \global \@minipagefalse
    \hb@xt@\hsize{\hfil\box\@tempboxa\hfil}%
  \fi
  \vskip\belowcaptionskip}

\renewenvironment{thebibliography}[1]{%
\begin{oldthebibliography}{#1}%
\small%
\raggedright%
\setlength{\itemsep}{5pt plus 0.2ex minus 0.05ex}%
}%
{%
\end{oldthebibliography}%
}

\begin{document}

%%%%%%%%%%%%%%%%%%±êÌâÒ³%%%%%%%%%%%%%%%%%%%%%%%%%%%%%&&&&&&&&&&&&&&&&&&&&&&

\title{\boldmath Scattering theory without large-distance asymptotics in arbitrary dimensions}

% more complex case: 4 authors, 3 institutions, 2 footnotes
\author[a]{Wen-Du Li}
\author[a,b,1]{and Wu-Sheng Dai}\note{daiwusheng@tju.edu.cn.}

% The "\note" macro will give a warning: "Ignoring empty anchor..."
% you can safely ignore it.

\affiliation[a]{Department of Physics, Tianjin University, Tianjin 300072, P.R. China}
\affiliation[b]{LiuHui Center for Applied Mathematics, Nankai University \& Tianjin University, Tianjin 300072, P.R. China}
%\affiliation[c]{DP School}

% e-mail addresses: one for each author, in the same order as the authors
%\emailAdd{Ccc@one.edu.cn}
%\emailAdd{second@asas.edu}
%\emailAdd{daiwusheng@tju.edu.cn}
%\emailAdd{fourth@one.univ}

%\title{\boldmath A title with some math: $x=1$}
%% %simple case: 2 authors, same institution
%% \author{A. Uthor}
%% \author{and A. Nother Author}
%% \affiliation{Institution,\\Address, Country}

% more complex case: 4 authors, 3 institutions, 2 footnotes
%\author[a,b,1]{F. Irst,\note{Corresponding author.}}
%\author[c]{S. Econd,}
%\author[a,2]{T. Hird\note{Also at Some University.}}
%\author[a,2]{and Fourth}

% The "\note" macro will give a warning: "Ignoring empty anchor..."
% you can safely ignore it.

%\affiliation[a]{One University,\\some-street, Country}
%\affiliation[b]{Another University,\\different-address, Country}
%\affiliation[c]{A School for Advanced Studies,\\some-location, Country}

% e-mail addresses: one for each author, in the same order as the authors

%\emailAdd{first@one.univ}
%\emailAdd{second@asas.edu}
%\emailAdd{third@one.univ}
%\emailAdd{fourth@one.univ}

%\date{date}

\abstract{
In conventional scattering theory, by large-distance asymptotics, at the cost
of losing the information of the distance between target and observer, one
imposes a large-distance asymptotics to achieve a scattering wave function
which can be represented explicitly by a scattering phase shift. In this
paper, without large-distance asymptotics, we establish an
arbitrary-dimensional scattering theory. Arbitrary-dimensional scattering wave
functions, scattering boundary conditions, cross sections, and phase shifts
are given without large-distance asymptotics. The importance of an
arbitrary-dimensional scattering theory is that the dimensional
renormalization procedure in quantum field theory needs an
arbitrary-dimensional result. Moreover, we give a discussion of one- and
two-dimensional scatterings.}

\maketitle
\flushbottom
%%%%%%%%%%%%%%%%%%±êÌâÒ³½áÊø%%%%%%%%%%%%%%%%%%%%%%%%%%%%%&&&&&&&&&&&&&&&&&&&

%%%%%%%%%%ÕýÎÄ¿ªÊ¼

\section{Introduction}

In conventional scattering theory, the information of the distance between
target and observer is lost due to large-distance asymptotics. By
large-distance asymptotics, in conventional three-dimensional scattering
theory, the\ solution of the radial Schr\"{o}dinger equation in the asymptotic
region and the\ scattering boundary condition are approximately represented as
\cite{ballentine1998quantum}%
\begin{align}
&  R_{l}\left(  r\right)  \overset{r\rightarrow\infty}{\sim}A_{l}\frac
{\sin\left(  kr-l\pi/2+\delta_{l}\right)  }{kr}, \label{yuanchangjingxiangjie}%
\\
&  \psi\left(  r,\theta\right)  \overset{r\rightarrow\infty}{\sim}\sum
_{l=0}^{\infty}\left(  2l+1\right)  i^{l}\frac{\sin\left(  kr-l\pi/2\right)
}{kr}P_{l}\left(  \cos\theta\right)  +f\left(  \theta\right)  \frac{e^{ikr}%
}{r}, \label{SRtiaojian}%
\end{align}
respectively, where $\delta_{l}$ is the scattering phase shift and $f\left(
\theta\right)  $ is the scattering amplitude.

Without large-distance asymptotics, in ref. \cite{liu2014scattering}, the
asymptotic solution (\ref{yuanchangjingxiangjie}) and the asymptotic
scattering boundary condition (\ref{SRtiaojian}) are replaced by the following
exact solutions:%
\begin{align}
R_{l}\left(  r\right)   &  =M_{l}\left(  -\frac{1}{ikr}\right)  \frac{A_{l}%
}{kr}\sin\left[  kr-\frac{l\pi}{2}+\delta_{l}+\Delta_{l}\left(  -\frac{1}%
{ikr}\right)  \right]  ,\label{Rlr}\\
\psi\left(  r,\theta\right)   &  =\sum_{l=0}^{\infty}\left(  2l+1\right)
i^{l}M_{l}\left(  -\frac{1}{ikr}\right)  \frac{1}{kr}\sin\left[  kr-\frac
{l\pi}{2}+\Delta_{l}\left(  -\frac{1}{ikr}\right)  \right]  P_{l}\left(
\cos\theta\right)  +f\left(  r,\theta\right)  \frac{e^{ikr}}{r},
\label{BCfrtheta}%
\end{align}
where $M_{l}\left(  x\right)  =\left\vert y_{l}\left(  x\right)  \right\vert $
and $\Delta_{l}\left(  x\right)  =\arg y_{l}\left(  x\right)  $ are the
modulus and argument of the Bessel polynomial $y_{l}\left(  x\right)  $
\cite{liu2014scattering}, respectively, and $P_{l}\left(  x\right)  $ is the
Legendre polynomial.

In this paper, we establish a rigorous scattering theory without
large-distance asymptotics in arbitrary dimensions.

In the following, we first rewrite the three-dimensional scattering theory
established in ref. \cite{liu2014scattering} in a new form which is convenient
to be generalized to arbitrary dimensions, and from which one can directly see
what happens after a scattering. We will show that, like that in
three-dimensional cases, the scattering phase shift is the only effect in
arbitrary-dimensional elastic scatterings.

Moreover, it will be shown that the scattering theory is different in odd and
even dimensions.

An arbitrary-dimensional scattering theory is important in quantum field
theory. For example, in scattering spectral method, to perform a dimensional
regularization procedure requires us to be able to carry out scattering theory
calculations in arbitrary dimensions
\cite{graham2009spectral,pang2012relation,li2015heat}. Moreover, two special
cases, one- and two-dimensional scatterings, are important both in theories
and experiments.

In section \ref{newform}, we rewrite the three-dimensional scattering theory
established in ref. \cite{liu2014scattering} in a new form. In section
\ref{wavefunction}, we give an exact $n$-dimensional scattering wave function
without large-distance asymptotics. In section \ref{boundarycondition}, we
construct a $n$-dimensional scattering boundary condition without
large-distance asymptotics. In section \ref{sin}, we rewrite the results given
in sections \ref{wavefunction} and \ref{boundarycondition} by sine functions,
which is the form in conventional scattering theory. In section
\ref{phaseshift}, we demonstrate how the scattering phase shift appears. In
section \ref{crosssection}, we give the $n$-dimensional scattering cross
section. In section \ref{123}, as examples, we discuss one-, two-, and
three-dimensional scatterings. In section \ref{n-asymptotics}, we demonstrate
how to take large-distance asymptotics of a $n$-dimensional scattering theory.
The conclusion is given in section \ref{Conclusions}.

\section{An alternative expression of three-dimensional scattering theory
without large-distance asymptotics \label{newform}}

In order to establish an arbitrary-dimensional scattering theory without
large-distance asymptotics, in this section, we rewrite the three-dimensional
scattering theory without large-distance asymptotics given in ref.
\cite{liu2014scattering} in a new form which is convenient to be generalized
to arbitrary dimensions.

\textit{For a three-dimensional scattering, without large-distance
asymptotics, the incident plane wave is}%
\begin{equation}
\psi^{in}\left(  r,\theta\right)  =\sum_{l=0}^{\infty}\left(  2l+1\right)
i^{l}\frac{1}{2}\left[  h_{l}^{\left(  2\right)  }\left(  kr\right)
+h_{l}^{\left(  1\right)  }\left(  kr\right)  \right]  P_{l}\left(  \cos
\theta\right)  ; \label{psiin3D}%
\end{equation}
\textit{after an elastic scattering, the wave function becomes}%
\begin{equation}
\psi\left(  r,\theta\right)  =\sum_{l=0}^{\infty}\left(  2l+1\right)
i^{l}\frac{1}{2}\left[  h_{l}^{\left(  2\right)  }\left(  kr\right)
+e^{2i\delta_{l}}h_{l}^{\left(  1\right)  }\left(  kr\right)  \right]
P_{l}\left(  \cos\theta\right)  , \label{psi3D}%
\end{equation}
\textit{where }$h_{\nu}^{\left(  1\right)  }\left(  z\right)  $\textit{ and
}$h_{\nu}^{\left(  2\right)  }\left(  z\right)  $\textit{ are the first and
second kind spherical Hankel functions.}

\textit{The scattering boundary condition is}%
\begin{equation}
\psi\left(  r,\theta\right)  =e^{ikr\cos\theta}+\sum_{l=0}^{\infty}%
a_{l}\left(  \theta\right)  h_{l}^{\left(  1\right)  }\left(  kr\right)  ,
\label{sbcal}%
\end{equation}
\textit{where}%
\begin{equation}
a_{l}\left(  \theta\right)  =\left(  2l+1\right)  i^{l}\frac{1}{2}\left(
e^{2i\delta_{l}}-1\right)  P_{l}\left(  \cos\theta\right)  . \label{altheta}%
\end{equation}
\textit{ }

\begin{proof}
The incident plane wave is
\begin{equation}
\psi^{in}\left(  r,\theta\right)  =e^{ikr\cos\theta}. \label{inplanewave}%
\end{equation}
Substituting the plane wave expansion $e^{ikr\cos\theta}=\sum_{l=0}^{\infty
}\left(  2l+1\right)  i^{l}j_{l}\left(  kr\right)  P_{l}\left(  \cos
\theta\right)  $ and $j_{l}\left(  z\right)  =\frac{1}{2}\left[
h_{l}^{\left(  1\right)  }\left(  z\right)  +h_{l}^{\left(  2\right)  }\left(
z\right)  \right]  $ \cite{liu2014scattering} into\ eq. (\ref{inplanewave})
proves eq. (\ref{psiin3D}) directly, where $j_{l}\left(  z\right)  $ is the
spherical Bessel function \cite{olver2010nist}.

Next we prove eqs. (\ref{psi3D}) and (\ref{sbcal}).

Without large-distance asymptotics, in ref. \cite{liu2014scattering}, we show
that the scattering boundary condition can be expressed as
\begin{equation}
\psi\left(  r,\theta\right)  =e^{ikr\cos\theta}+f\left(  r,\theta\right)
\frac{e^{ikr}}{r} \label{Psifrtheta}%
\end{equation}
with%
\begin{equation}
f\left(  r,\theta\right)  =\frac{1}{2ik}\sum_{l=0}^{\infty}\left(
2l+1\right)  \left(  e^{2i\delta_{l}}-1\right)  P_{l}\left(  \cos
\theta\right)  y_{l}\left(  -\frac{1}{ikr}\right)  . \label{frtheta}%
\end{equation}
By eq. (\ref{altheta}), we can rewrite $f\left(  r,\theta\right)  $ as
\begin{equation}
f\left(  r,\theta\right)  =\frac{1}{k}\sum_{l=0}^{\infty}a_{l}\left(
\theta\right)  \left(  -i\right)  ^{l+1}y_{l}\left(  -\frac{1}{ikr}\right)  .
\label{frthtatht}%
\end{equation}
Substituting eq. (\ref{frthtatht}) into eq. (\ref{Psifrtheta}) and using
$h_{l}^{\left(  1\right)  }\left(  z\right)  =\left(  -i\right)  ^{l+1}\left(
e^{iz}/z\right)  y_{l}\left(  i/z\right)  $ \cite{liu2014scattering}, we
arrive at eq. (\ref{sbcal}).

Finally, substituting eq. (\ref{psiin3D}) into (\ref{sbcal}) and using
(\ref{altheta}) prove eq. (\ref{psi3D}).
\end{proof}

By large-distance asymptotics, in conventional scattering theory, it is proved
that the phase shift is the only effect after an elastic scattering and all
information of an elastic scattering is embedded in a scattering phase shift
\cite{ballentine1998quantum}.

Without large-distance asymptotics, it is proved by comparing eqs.
(\ref{psiin3D}) and (\ref{psi3D}) that, the only effect after an elastic
scattering is still a phase shift on the outgoing wave function: the incoming
part, represented by $h_{l}^{\left(  2\right)  }\left(  kr\right)  $, does not
change anymore; the outgoing part, represented by $h_{l}^{\left(  1\right)
}\left(  kr\right)  $, changes a phase factor $e^{2i\delta_{l}}$.

Naturally, when taking large-distance asymptotics, the above result will
reduce to conventional scattering theory: $\sum_{l=0}^{\infty}a_{l}\left(
\theta\right)  h_{l}^{\left(  1\right)  }\left(  kr\right)
\overset{r\rightarrow\infty}{\sim}f\left(  \theta\right)  e^{ikr}/r$ with the
scattering amplitude $f\left(  \theta\right)  =\sum_{l=0}^{\infty}a_{l}\left(
\theta\right)  /\left(  i^{l+1}k\right)  $.

\section{$n$-dimensional scattering wave function \label{wavefunction}\ }

For a $n$-dimensional scattering, the radial wave equation with a spherical
potential reads \cite{graham2009spectral}
\begin{equation}
\left[  \frac{d^{2}}{dr^{2}}+\frac{n-1}{r}\frac{d}{dr}+k^{2}-\frac{l\left(
l+n-2\right)  }{r^{2}}-V\left(  r\right)  \right]  R_{l}\left(  r\right)  =0.
\label{radialeq}%
\end{equation}

The solution of the asymptotic equation of the radial equation (\ref{radialeq}%
), i.e., eq. (\ref{radialeq}) with $V\left(  r\right)  =0$, can be solved
exactly:%
\begin{equation}
R_{l}\left(  r\right)  =C_{l}\frac{h_{l+\left(  n-3\right)  /2}^{\left(
2\right)  }\left(  kr\right)  }{r^{\left(  n-3\right)  /2}}+D_{l}%
\frac{h_{l+\left(  n-3\right)  /2}^{\left(  1\right)  }\left(  kr\right)
}{r^{\left(  n-3\right)  /2}}. \label{RlrexactCD}%
\end{equation}
It should be noted that in conventional scattering theory the exact solution
of the asymptotic equation (\ref{radialeq}) is approximated by an asymptotic
solution of the asymptotic equation, like eq. (\ref{yuanchangjingxiangjie}).

The $n$-dimensional wave function can be expressed as%
\[
\psi\left(  r,\theta\right)  =\sum_{l=0}^{\infty}R_{l}\left(  r\right)
C_{l}^{n/2-1}\left(  \cos\theta\right)  ,
\]
where $C_{l}^{\lambda}\left(  z\right)  $ is the Gegenbauer polynomial, a
generalization of the Legendre polynomial \cite{olver2010nist}. Then, by eq.
(\ref{RlrexactCD}), we arrive at
\begin{equation}
\psi\left(  r,\theta\right)  =\sum_{l=0}^{\infty}C_{l}\left[  \frac
{h_{l+\left(  n-3\right)  /2}^{\left(  2\right)  }\left(  kr\right)
}{r^{\left(  n-3\right)  /2}}+e^{2i\delta_{l}}\frac{h_{l+\left(  n-3\right)
/2}^{\left(  1\right)  }\left(  kr\right)  }{r^{\left(  n-3\right)  /2}%
}\right]  C_{l}^{n/2-1}\left(  \cos\theta\right)  , \label{wf2}%
\end{equation}
where $e^{2i\delta_{l}}=D_{l}/C_{l}$ defines the phase shift
\cite{liu2014scattering}.

\section{$n$-dimensional scattering boundary condition
\label{boundarycondition}}

A scattering is determined by the Schr\"{o}dinger equation with a scattering
boundary condition. In conventional scattering theory, the scattering boundary
condition is the Sommerfeld radiation condition which is constructed under
large-distance asymptotics. In our preceding work \cite{liu2014scattering},
without large-distance asymptotics, instead of the Sommerfeld radiation
condition, we construct a scattering boundary condition, eq. (\ref{BCfrtheta})
or, equivalently, eq. (\ref{sbcal}), which preserves the information of the
distance between the target and observer.

\subsection{Scattering boundary condition}

In the following, without large-distance asymptotics, we construct the
$n$-dimensional scattering boundary condition.

Generally speaking, a scattering boundary condition is a wave function at an
asymptotic distance, consisting of two parts: the incident wave $\psi^{in}$
and the scattering wave $\psi^{sc}$, i.e., $\psi=\psi^{in}+\psi^{sc}$. In
three-dimensional conventional scattering theory, $\psi^{sc}$ is chosen as
being in proportion to $e^{ikr}/r$, since the asymptotics of the scattering
wave function is $R_{l}\overset{r\rightarrow\infty}{\sim}e^{\pm ikr}/r$ and
only the outgoing wave $R_{l}\overset{r\rightarrow\infty}{\sim}e^{ikr}/r$
remains in the scattering wave function when $r\rightarrow\infty$
\cite{joachain1975quantum}. Without the asymptotic approximation, as shown in
eq. (\ref{RlrexactCD}), the solution is $h_{l+\left(  n-3\right)  /2}^{\left(
1,2\right)  }\left(  kr\right)  /r^{\left(  n-3\right)  /2}$ and only the
outgoing wave $h_{l+\left(  n-3\right)  /2}^{\left(  1\right)  }\left(
kr\right)  /r^{\left(  n-3\right)  /2}$ remains in the scattering wave
function. To retrieve the information of the distance, we construct the
scattering boundary condition by $h_{l+\left(  n-3\right)  /2}^{\left(
1\right)  }\left(  kr\right)  /r^{\left(  n-3\right)  /2}$ rather than its
asymptotics $e^{ikr}/r$.

To generalize the three-dimensional scattering boundary condition to $n$
dimensions, we replace the three-dimensional outgoing wave $h_{l}^{\left(
1\right)  }\left(  kr\right)  $ in eq. (\ref{sbcal}) with $n$-dimensional
outgoing wave $h_{l+\left(  n-3\right)  /2}^{\left(  1\right)  }\left(
kr\right)  /r^{\left(  n-3\right)  /2}$:%
\begin{equation}
\psi\left(  r,\theta\right)  =e^{ikr\cos\theta}+\sum_{l=0}^{\infty}%
a_{l}\left(  \theta\right)  \frac{h_{l+\left(  n-3\right)  /2}^{\left(
1\right)  }\left(  kr\right)  }{r^{\left(  n-3\right)  /2}}.
\label{biantiaojian}%
\end{equation}
The expression of $a_{l}\left(  \theta\right)  $ will be given in the following.

\subsection{$a_{l}\left(  \theta\right)  $}

In a scattering theory without large-distance asymptotics, $a_{l}\left(
\theta\right)  $ plays the role of the partial wave scattering amplitude in
conventional scattering theory, and the information of the scattering is
embedded in $a_{l}\left(  \theta\right)  $. In this section, we calculate
$a_{l}\left(  \theta\right)  $ in $n$ dimensions.

By using the $n$-dimensional plane wave expansion \cite{olver2010nist}
\begin{equation}
e^{ikr\cos\theta}=\frac{\Gamma\left(  n/2-1\right)  }{\sqrt{\pi}\left(
k/2\right)  ^{\left(  n-3\right)  /2}}\sum_{l=0}^{\infty}\left(
2l+n-2\right)  i^{l}\frac{j_{l+\left(  n-3\right)  /2}\left(  kr\right)
}{r^{\left(  n-3\right)  /2}}C_{l}^{n/2-1}\left(  \cos\theta\right)
\label{ndpmb}%
\end{equation}
and $j_{l}\left(  z\right)  =\frac{1}{2}\left[  h_{l}^{\left(  1\right)
}\left(  z\right)  +h_{l}^{\left(  2\right)  }\left(  z\right)  \right]  $, we
can rewrite the scattering boundary condition (\ref{biantiaojian}) as
\begin{align}
\psi\left(  r,\theta\right)   &  =\sum_{l=0}^{\infty}\left[  \frac
{\Gamma\left(  n/2-1\right)  }{\sqrt{\pi}\left(  k/2\right)  ^{\left(
n-3\right)  /2}}\left(  2l+n-2\right)  i^{l}\frac{1}{2}C_{l}^{n/2-1}\left(
\cos\theta\right)  +a_{l}\left(  \theta\right)  \right]  \frac{h_{l+\left(
n-3\right)  /2}^{\left(  1\right)  }\left(  kr\right)  }{r^{\left(
n-3\right)  /2}}\nonumber\\
&  +\frac{\Gamma\left(  n/2-1\right)  }{\sqrt{\pi}\left(  k/2\right)
^{\left(  n-3\right)  /2}}\sum_{l=0}^{\infty}\left(  2l+n-2\right)  i^{l}%
\frac{1}{2}\frac{h_{l+\left(  n-3\right)  /2}^{\left(  2\right)  }\left(
kr\right)  }{r^{\left(  n-3\right)  /2}}C_{l}^{n/2-1}\left(  \cos
\theta\right)  . \label{wf1}%
\end{align}

Then $a_{l}\left(  \theta\right)  $ can be achieved immediately by equating
the coefficients in eqs. (\ref{wf1}) and (\ref{wf2}):
\begin{align}
\frac{\Gamma\left(  n/2-1\right)  }{\sqrt{\pi}\left(  k/2\right)  ^{\left(
n-3\right)  /2}}\left(  2l+n-2\right)  i^{l}\frac{1}{2}C_{l}^{n/2-1}\left(
\cos\theta\right)  +a_{l}\left(  \theta\right)   &  =C_{l}e^{2i\delta_{l}%
}C_{l}^{n/2-1}\left(  \cos\theta\right)  ,\label{aa1}\\
\frac{\Gamma\left(  n/2-1\right)  }{\sqrt{\pi}\left(  k/2\right)  ^{\left(
n-3\right)  /2}}\left(  2l+n-2\right)  i^{l}\frac{1}{2}  &  =C_{l}.
\label{aa2}%
\end{align}
Substituting eq. (\ref{aa2}) into eq. (\ref{aa1}) gives
\begin{equation}
a_{l}\left(  \theta\right)  =\frac{\Gamma\left(  n/2-1\right)  }{\sqrt{\pi
}\left(  k/2\right)  ^{\left(  n-3\right)  /2}}\left(  2l+n-2\right)
i^{l}\frac{1}{2}\left(  e^{2i\delta_{l}}-1\right)  C_{l}^{n/2-1}\left(
\cos\theta\right)  . \label{al}%
\end{equation}

It should be emphasized that $n=2$ is a removable singularity\ of
$a_{l}\left(  \theta\right)  $, which will be discussed in Sec. \ref{2D}.

\section{Representing scattering wave function by sine function \label{sin}}

In conventional scattering theory, the scattering wave function is
approximately expressed by a sine function. In Ref. \cite{liu2014scattering},
we show that the scattering wave function, in fact, can be exactly expressed
by a sine function. In this section, we represent the $n$-dimensional
scattering wave function by a sine function exactly.

\subsection{Radial wave function}

In order to represent the scattering wave function by a sine function, we
first rewrite the spherical Hankel function as%

\begin{align}
h_{\nu}^{\left(  1\right)  }\left(  z\right)   &  =e^{i\left(  z-\nu
\pi/2\right)  }\frac{1}{iz}\mathcal{Y}_{\nu}\left(  -\frac{1}{iz}\right)
,\label{sph1}\\
h_{\nu}^{\left(  2\right)  }\left(  z\right)   &  =-e^{-i\left(  z-\nu
\pi/2\right)  }\frac{1}{iz}\mathcal{Y}_{\nu}\left(  \frac{1}{iz}\right)  .
\label{sph2}%
\end{align}
Here we introduce%
\begin{equation}
\mathcal{Y}_{\nu}\left(  z\right)  =\left(  \frac{2}{z}\right)  ^{\nu
+1}U\left(  \nu+1,2\left(  \nu+1\right)  ,\frac{2}{z}\right)  , \label{Yvz}%
\end{equation}
where $U\left(  a,b;z\right)  $ is the Tricomi confluent hypergeometric
function \cite{olver2010nist}. Notice that $\mathcal{Y}_{\nu}\left(  z\right)
$ recovers the Bessel polynomial in odd dimensions.

The radial wave function (\ref{RlrexactCD}) then can be expressed as%
\begin{align}
R_{l}\left(  r\right)   &  =C_{l}\left\{  -\frac{e^{-i\left\{  kr-\left[
l+\left(  n-3\right)  /2\right]  \left(  \pi/2\right)  \right\}  }}{ikr}%
\frac{1}{r^{\left(  n-3\right)  /2}}\mathcal{Y}_{l+\left(  n-3\right)
/2}\left(  \frac{1}{ikr}\right)  \right. \nonumber\\
&  +\left.  e^{2i\delta_{l}}\frac{e^{i\left\{  kr-\left[  l+\left(
n-3\right)  /2\right]  \left(  \pi/2\right)  \right\}  }}{ikr}\frac
{1}{r^{\left(  n-3\right)  /2}}\mathcal{Y}_{l+\left(  n-3\right)  /2}\left(
-\frac{1}{ikr}\right)  \right\}  ; \label{RlrYY}%
\end{align}
notice that here $e^{2i\delta_{l}}=D_{l}/C_{l}$. Then the radial wave function
(\ref{RlrYY}) can be represented by a sine function:%
\begin{equation}
R_{l}\left(  r\right)  =M_{l}\left(  -\frac{1}{ikr}\right)  \frac{A_{l}}%
{kr}\frac{1}{r^{\left(  n-3\right)  /2}}\sin\left[  kr-\left(  l+\frac{n-3}%
{2}\right)  \frac{\pi}{2}+\delta_{l}+\Delta_{l}\left(  -\frac{1}{ikr}\right)
\right]  ,
\end{equation}
where $A_{l}=2\sqrt{C_{l}D_{l}}$ and $M_{l}=\left\vert \mathcal{Y}_{l+\left(
n-3\right)  /2}\left(  -\frac{1}{ikr}\right)  \right\vert $ and $\Delta
_{l}=\arg\mathcal{Y}_{l+\left(  n-3\right)  /2}\left(  -\frac{1}{ikr}\right)
$ are the modulus and argument of $\mathcal{Y}_{l+\left(  n-3\right)
/2}\left(  -\frac{1}{ikr}\right)  $, respectively.

When employing large-distance asymptotics, the radial wave function becomes%
\begin{equation}
R_{l}\left(  r\right)  \overset{r\rightarrow\infty}{\sim}\frac{A_{l}}{kr}%
\frac{1}{r^{\left(  n-3\right)  /2}}\sin\left[  kr-\left(  l+\frac{n-3}%
{2}\right)  \frac{\pi}{2}+\delta_{l}\right]  ,
\end{equation}
where asymptotics $U\left(  a,b;z\right)  \overset{r\rightarrow\infty}{\sim
}1/z^{a}$ \cite{olver2010nist} and eq. (\ref{hjianjin}) which will be proved
in section \ref{n-asymptotics} are used.

\subsection{$\mathcal{Y}_{\nu}\left(  z\right)  $ in odd and even dimensions}

As will be shown in the following, the function $\mathcal{Y}_{\nu}\left(
z\right)  $ defined by eq. (\ref{Yvz}) is different in odd and even
dimensions: in odd dimensions $\mathcal{Y}_{\nu}\left(  z\right)  $ is a
polynomial and in even dimensions $\mathcal{Y}_{\nu}\left(  z\right)  $ is an
infinite series.

The Tricomi confluent hypergeometric function $U\left(  a,b;z\right)  $, when
$b=m+1$ ($m=0,1,2,...$), can be expanded as \cite{olver2010nist}%
\begin{align}
U\left(  a,m+1,z\right)   &  =\frac{\left(  -1\right)  ^{m+1}}{m!\Gamma\left(
a-m\right)  }\sum_{k=0}^{\infty}\frac{\left(  a\right)  _{k}}{\left(
m+1\right)  _{k}k!}z^{k}\left[  \ln z+\psi\left(  a+k\right)  -\psi\left(
1+k\right)  -\psi\left(  m+k-1\right)  \right] \nonumber\\
&  +\frac{1}{\Gamma\left(  a\right)  }\sum_{k=1}^{m}\frac{\left(  k-1\right)
!\left(  1-a+k\right)  _{m-k}}{\left(  m-k\right)  !}\frac{1}{z^{k}},
\end{align}
where $\left(  \alpha\right)  _{n}=\alpha\left(  \alpha+1\right)
\ldots\left(  \alpha+n-1\right)  =\Gamma\left(  \alpha+n\right)
/\Gamma\left(  \alpha\right)  $ is the Pochhammer symbol and $\psi\left(
z\right)  =\Gamma^{\prime}\left(  z\right)  /\Gamma\left(  z\right)  $ is the
digamma function \cite{olver2010nist}.

Then by eq. (\ref{Yvz}), for $n\geq2$, we have%
\begin{align}
\mathcal{Y}_{l+\left(  n-3\right)  /2}\left(  z\right)   &  =\left(  \frac
{2}{z}\right)  ^{l+\left(  n-1\right)  /2}\sum_{j=0}^{2l+n-3}\frac
{\Gamma\left(  j+1\right)  }{\Gamma\left(  -l-n/2+5/2+j\right)  \Gamma\left(
2l+n-2-j\right)  }\left(  \frac{z}{2}\right)  ^{j+1}\nonumber\\
&  +\left(  \frac{2}{z}\right)  ^{l+\left(  n-1\right)  /2}\frac{\left(
-1\right)  ^{2l+n-1}}{\left(  2l+n-2\right)  !\Gamma\left(  -l-\left(
n-3\right)  /2\right)  }\sum_{j=0}^{\infty}\frac{\left(  l+\left(  n-1\right)
/2\right)  _{j}}{\left(  2l+n-1\right)  _{j}j!}\left(  \frac{2}{z}\right)
^{j}\nonumber\\
&  \times\left[  \ln\left(  \frac{2}{z}\right)  +\psi\left(  l+\frac{n-1}%
{2}+j\right)  -\psi\left(  1+j\right)  -\psi\left(  2l+n-3+j\right)  \right]
. \label{Y}%
\end{align}

For odd-dimensional cases ($n\neq1$), i.e., $n=3,5,...$, the second term in
eq. (\ref{Y}) equals zero because $1/\Gamma\left(  -l-\left(  n-3\right)
/2\right)  =0$. Moreover, because $1/\Gamma\left(  -l-n/2+5/2+j\right)  =0$
when $-l-n/2+5/2+j=0,-1,-2,...$, the summation of $j$ in fact begins with
$j=l+\left(  n-3\right)  /2$ rather than $j=0$. Therefore, in odd dimensions
($n\neq1$), $\mathcal{Y}_{l+\left(  n-3\right)  /2}\left(  z\right)  $ is in
fact a polynomial,%
\begin{equation}
\mathcal{Y}_{l+\left(  n-3\right)  /2}\left(  z\right)  =\left(  \frac{2}%
{z}\right)  ^{l+\left(  n-1\right)  /2}\sum_{j=l+\left(  n-3\right)
/2}^{2l+n-3}\frac{\Gamma\left(  j+1\right)  }{\Gamma\left(
-l-n/2+5/2+j\right)  \Gamma\left(  2l+n-2-j\right)  }\left(  \frac{z}%
{2}\right)  ^{j+1}.
\end{equation}
This result can be rewritten as%
\begin{equation}
\mathcal{Y}_{\nu}\left(  z\right)  =\sum_{j=0}^{\nu}\frac{\left(
\nu+j\right)  !}{j!\left(  \nu-j\right)  !}\left(  \frac{z}{2}\right)
^{j}=y_{\nu}\left(  z\right)  ,
\end{equation}
where $y_{\nu}\left(  z\right)  $ is just the Bessel polynomial. That is to
say, $\mathcal{Y}_{\nu}\left(  z\right)  $ recovers the Bessel polynomial
$y_{l}\left(  z\right)  $ in odd dimensions.

On the contrary, for even-dimensional cases, $n$ is an even number and
$\nu=l+\left(  n-3\right)  /2$ is a half-integer; as a result, $\mathcal{Y}%
_{\nu}\left(  z\right)  $ is an infinite series rather than a polynomial.

\section{Wave functions before and after a scattering: phase shifts
\label{phaseshift}}

In this section, similar to eqs. (\ref{psiin3D})\ and (\ref{psi3D}), we write
out the $n$-dimensional\ wave functions before and after a scattering. The
phase shift is the only effect in an elastic scattering process, i.e., all
information of an elastic scattering process is embedded in a scattering phase
shift \cite{ballentine1998quantum}. The result given below will show how a
scattering phase shift appears.

The $n$-dimensional\ incident plane wave, by eq. (\ref{ndpmb}), can be
expressed as
\begin{equation}
\psi^{in}\left(  r,\theta\right)  =\frac{\Gamma\left(  n/2-1\right)  }%
{\sqrt{\pi}\left(  k/2\right)  ^{\left(  n-3\right)  /2}}\sum_{l=0}^{\infty
}\left(  2l+n-2\right)  i^{l}\frac{1}{2}\left[  \frac{h_{l+\left(  n-3\right)
/2}^{\left(  2\right)  }\left(  kr\right)  }{r^{\left(  n-3\right)  /2}}%
+\frac{h_{l+\left(  n-3\right)  /2}^{\left(  1\right)  }\left(  kr\right)
}{r^{\left(  n-3\right)  /2}}\right]  C_{l}^{n/2-1}\left(  \cos\theta\right)
. \label{bef}%
\end{equation}
After an elastic scattering, the wave function, by eqs. (\ref{biantiaojian})
and (\ref{al}), becomes
\begin{equation}
\psi\left(  r,\theta\right)  =\frac{\Gamma\left(  n/2-1\right)  }{\sqrt{\pi
}\left(  k/2\right)  ^{\left(  n-3\right)  /2}}\sum_{l=0}^{\infty}\left(
2l+n-2\right)  i^{l}\frac{1}{2}\left[  \frac{h_{l+\left(  n-3\right)
/2}^{\left(  2\right)  }\left(  kr\right)  }{r^{\left(  n-3\right)  /2}%
}+e^{2i\delta_{l}}\frac{h_{l+\left(  n-3\right)  /2}^{\left(  1\right)
}\left(  kr\right)  }{r^{\left(  n-3\right)  /2}}\right]  C_{l}^{n/2-1}\left(
\cos\theta\right)  . \label{aft}%
\end{equation}

Comparing the wave functions before and after a scattering process, eqs.
(\ref{bef}) and (\ref{aft}), we can see that after a scattering, the incoming
part which is represented by $h_{l+\left(  n-3\right)  /2}^{\left(  2\right)
}\left(  kr\right)  /r^{\left(  n-3\right)  /2}$, does not change anymore,
while a phase factor $e^{2i\delta_{l}}$ appears in the outgoing part which is
represented by $h_{l+\left(  n-3\right)  /2}^{\left(  1\right)  }\left(
kr\right)  /r^{\left(  n-3\right)  /2}$. This reveals that the only effect
after an elastic scattering is a phase shift on the outgoing wave function.

\section{$n$-dimensional differential scattering cross section
\label{crosssection}}

Without large-distance asymptotics, the $n$-dimensional differential
scattering cross section is%
\begin{equation}
\frac{d\sigma}{d\Omega}=\frac{\mathbf{j}^{sc}\cdot d\mathbf{S}}{j^{in}}%
=\frac{\left\vert \mathbf{j}^{sc}\right\vert }{j^{in}\cos\gamma}r^{n-1}%
=\frac{\sqrt{\left(  j_{r}^{sc}\right)  ^{2}+\left(  j_{\theta}^{sc}\right)
^{2}}}{j^{in}\cos\gamma}r^{n-1},\label{weifenjiemian}%
\end{equation}
where $d\mathbf{S}=\mathbf{\hat{n}}dS=\mathbf{\hat{n}}r^{n-1}d\Omega$ with the
$n$-dimensional solid angle\newline $d\Omega=\sin^{n-2}\theta_{1}\sin
^{n-3}\theta_{2}...\sin\theta_{n-2}d\theta_{1}d\theta_{2}...d\theta_{n-2}%
d\phi$ and $\gamma$ is the angle between $\mathbf{j}^{sc}$ and $\mathbf{\hat
{r}}$,
\begin{equation}
\tan\gamma=\frac{j_{\theta}^{sc}}{j_{r}^{sc}}.
\end{equation}
The $n$-dimensional differential scattering cross section (\ref{weifenjiemian}%
) can\ be rewritten as
\begin{equation}
\frac{d\sigma}{d\Omega}=\frac{j_{r}^{sc}}{j^{in}}\left[  1+\left(
\frac{j_{\theta}^{sc}}{j_{r}^{sc}}\right)  ^{2}\right]  r^{n-1}.
\end{equation}

The leading contribution of the $n$-dimensional differential scattering cross
section is%
\begin{equation}
\frac{d\sigma}{d\Omega}=\frac{j_{r}^{sc}}{j^{in}}r^{n-1}=\frac{1}%
{k}\operatorname{Im}\left(  \psi^{sc\ast}\frac{\partial}{\partial r}\psi
^{sc}\right)  r^{n-1}. \label{weifenjiemian1}%
\end{equation}
Substituting $\psi^{sc}=\psi-\psi^{in}=\sum_{l=0}^{\infty}a_{l}\left(
\theta\right)  h_{l+\left(  n-3\right)  /2}^{\left(  1\right)  }\left(
kr\right)  /r^{\left(  n-3\right)  /2}$ into eq. (\ref{weifenjiemian1}), we
have
\begin{equation}
\frac{d\sigma}{d\Omega}=r^{n-1}\frac{1}{2ik}\sum_{l=0}^{\infty}\sum
_{l^{\prime}=0}^{\infty}a_{l}^{\ast}\left(  \theta\right)  a_{l^{\prime}%
}\left(  \theta\right)  W_{r}\left[  \frac{h_{l+\left(  n-3\right)
/2}^{\left(  2\right)  }\left(  kr\right)  }{r^{\left(  n-3\right)  /2}}%
,\frac{h_{l^{\prime}+\left(  n-3\right)  /2}^{\left(  1\right)  }\left(
kr\right)  }{r^{\left(  n-3\right)  /2}}\right]  , \label{sigma}%
\end{equation}
where $W\left[  f\left(  r\right)  ,g\left(  r\right)  \right]  =f\left(
r\right)  \frac{d}{dr}g\left(  r\right)  -g\left(  r\right)  \frac{d}%
{dr}f\left(  r\right)  $ is the Wronskian determinant.

\section{One-, two-, and three-dimensional scatterings \label{123}}

In this section, we discuss one-, two-, and three-dimensional scatterings,
respectively. These three kinds of scattering can occur in real physical systems.

\subsection{One-dimensional scattering}

For a one-dimensional scattering, $n=1$, by eq. (\ref{al}), we have%
\begin{equation}
a_{l}\left(  \theta\right)  =-k\left(  2l-1\right)  i^{l}\frac{1}{2}\left(
e^{2i\delta_{l}}-1\right)  C_{l}^{-1/2}\left(  \cos\theta\right)  .
\end{equation}
In the one-dimensional case, $\theta$ can only take two possible values, $0$
and $\pi$. Therefore,%
\begin{align}
C_{0}^{-1/2}\left(  1\right)   &  =C_{0}^{-1/2}\left(  -1\right)
=C_{1}^{-1/2}\left(  -1\right)  =-C_{1}^{-1/2}\left(  1\right)  =1,\nonumber\\
C_{l}^{-1/2}\left(  \cos\theta\right)   &  =0,\text{\ \ }l\neq0,1.
\end{align}
Thus we have%
\begin{align}
a_{0}\left(  0\right)   &  =a_{0}\left(  \pi\right)  =-\frac{1}{2}k\left(
e^{2i\delta_{0}}-1\right)  ,\text{\ }\nonumber\\
\text{\ }a_{1}\left(  0\right)   &  =-a_{1}\left(  \pi\right)  =-\frac{1}%
{2}ik\left(  e^{2i\delta_{1}}-1\right)  ,\nonumber\\
a_{l}\left(  0\right)   &  =a_{l}\left(  \pi\right)  =0,\text{ \ \ }l\neq0,1.
\end{align}
From eq. (\ref{sigma}), we obtain the differential scattering cross section at
$\theta=0$ and $\theta=\pi$:%
\begin{align}
\sigma\left(  0\right)   &  =\sin^{2}\delta_{0}+\sin^{2}\delta_{1}%
+2\cos\left(  \delta_{0}-\delta_{1}\right)  \sin\delta_{0}\sin\delta_{1},\\
\sigma\left(  \pi\right)   &  =\sin^{2}\delta_{0}+\sin^{2}\delta_{1}%
-2\cos\left(  \delta_{0}-\delta_{1}\right)  \sin\delta_{0}\sin\delta_{1}.
\end{align}

In a one-dimensional scattering, we are interested in transmissivity $T$ and
reflectivity $R$:%

\begin{align}
T  &  =\frac{\sigma\left(  0\right)  }{\sigma\left(  0\right)  +\sigma\left(
\pi\right)  }=\frac{1}{2}+\frac{\cos\left(  \delta_{0}-\delta_{1}\right)
\sin\delta_{0}\sin\delta_{1}}{\sin^{2}\delta_{0}+\sin^{2}\delta_{1}},\\
R  &  =\frac{\sigma\left(  \pi\right)  }{\sigma\left(  0\right)
+\sigma\left(  \pi\right)  }=\frac{1}{2}-\frac{\cos\left(  \delta_{0}%
-\delta_{1}\right)  \sin\delta_{0}\sin\delta_{1}}{\sin^{2}\delta_{0}+\sin
^{2}\delta_{1}}.
\end{align}

It can be seen that in one dimension the scattering result is independent of
the distance $r$.

\subsection{Two-dimensional scattering \label{2D}}

In a two-dimensional scattering, we encounter a singularity in $a_{l}\left(
\theta\right)  $ given by eq. (\ref{al}). We will show that, however, $n=2$ is
a removable singularity.

We can see from eq. (\ref{al}) that $n=2$ is a singularity of the gamma
function $\Gamma\left(  n/2-1\right)  $, but, meanwhile, $n=2$ is also a zero
of the Gegenbauer polynomial $C_{l}^{n/2-1}\left(  \cos\theta\right)  $. This
makes $n=2$ a removable singularity.

When $n=2$, $a_{l}\left(  \theta\right)  $ given by eq. (\ref{al}) reduces to%
\begin{equation}
a_{l}\left(  \theta\right)  =\operatorname*{Deg}\left(  l\right)  \frac{1}%
{2}i^{l}\sqrt{\frac{2k}{\pi}}\left(  e^{2i\delta_{l}}-1\right)  \cos\left(
l\theta\right)  ,
\end{equation}
where $\operatorname*{Deg}\left(  l\right)  $ is the degeneracy,%
\begin{align}
\operatorname*{Deg}\left(  l\right)   &  =1,\text{ \ \ }l=0,\\
\operatorname*{Deg}\left(  l\right)   &  =2,\text{ \ \ }l\neq0.
\end{align}
The differential scattering cross section can be obtained by eq. (\ref{sigma})
with $n=2$:%
\begin{align}
\frac{d\sigma}{d\Omega}  &  =r\frac{1}{2ik}\sum_{l=0}^{\infty}\sum_{l^{\prime
}=0}^{\infty}a_{l}^{\ast}\left(  \theta\right)  a_{l^{\prime}}\left(
\theta\right)  W_{r}\left[  \frac{h_{l-1/2}^{\left(  2\right)  }\left(
kr\right)  }{r^{-1/2}},\frac{h_{l^{\prime}-1/2}^{\left(  1\right)  }\left(
kr\right)  }{r^{-1/2}}\right] \nonumber\\
&  =\frac{kr}{2\pi}\sum_{l=0}^{\infty}\sum_{l^{\prime}=0}^{\infty
}\operatorname*{Deg}\left(  l\right)  \operatorname*{Deg}\left(  l^{\prime
}\right)  \left(  -i\right)  ^{l}i^{l^{\prime}}\left(  e^{-2i\delta_{l}%
}-1\right)  \left(  e^{2i\delta_{l^{\prime}}}-1\right)  \cos\left(
l\theta\right)  \cos\left(  l^{\prime}\theta\right) \nonumber\\
&  \times\frac{1}{2ik}W_{r}\left[  \sqrt{r}h_{l-1/2}^{\left(  2\right)
}\left(  kr\right)  ,\sqrt{r}h_{l^{\prime}-1/2}^{\left(  1\right)  }\left(
kr\right)  \right]  .
\end{align}

Performing large-distance asymptotics gives
\begin{align}
\frac{d\sigma}{d\Omega}  &  \sim\frac{1}{2\pi k}\sum_{l=0}^{\infty}%
\sum_{l^{\prime}=0}^{\infty}\operatorname*{Deg}\left(  l\right)
\operatorname*{Deg}\left(  l^{\prime}\right)  \left(  e^{-2i\delta_{l}%
}-1\right)  \left(  e^{2i\delta_{l^{\prime}}}-1\right)  \cos\left(
l\theta\right)  \cos\left(  l^{\prime}\theta\right) \\
&  =\left\vert f\left(  \theta\right)  \right\vert ^{2}.
\end{align}

\subsection{Three-dimensional scattering}

The three-dimensional result can be obtained directly by setting the dimension
$n=3$ in the above result.

Eq. (\ref{al}) with $n=3$, by the relation $C_{l}^{1/2}\left(  \cos
\theta\right)  =P_{l}\left(  \cos\theta\right)  $ \cite{olver2010nist}, gives
eq. (\ref{altheta}). The differential scattering cross section can then be
obtained by eq. (\ref{sigma}):%
\begin{equation}
\frac{d\sigma}{d\Omega}=r^{2}\sum_{l=0}^{\infty}\sum_{l^{\prime}=0}^{\infty
}a_{l}^{\ast}\left(  \theta\right)  a_{l^{\prime}}\left(  \theta\right)
\frac{W_{r}\left[  h_{l}^{\left(  2\right)  }\left(  kr\right)  ,h_{l^{\prime
}}^{\left(  1\right)  }\left(  kr\right)  \right]  }{2ik}.
\end{equation}
This result agrees with the result given by Ref. \cite{liu2014scattering}.

\section{$n$-dimensional scattering with large-distance asymptotics
\label{n-asymptotics}}

In the above, we obtain a $n$-dimensional scattering theory without
large-distance asymptotics, which contains the information of the distance
between target and observer. In this section, we demonstrate how this result
reduces to the conventional scattering result when taking $r\rightarrow\infty$ asymptotics.

Using \cite{olver2010nist}%
\begin{equation}
h_{\nu}^{\left(  1,2\right)  }\left(  z\right)  =\sqrt{\frac{\pi}{2z}}H_{\nu
}^{\left(  1,2\right)  }\left(  z\right)
\end{equation}
and%
\begin{equation}
H_{\nu}^{\left(  1,2\right)  }\left(  z\right)  =\mp\frac{2}{\sqrt{\pi}%
}ie^{\mp i\nu\pi}\left(  2z\right)  ^{\nu}e^{\pm iz}U\left(  \nu+\frac{1}%
{2},2\nu+1,\mp2iz\right)  ,
\end{equation}
where $H_{\nu}^{\left(  1\right)  }\left(  z\right)  $ and $H_{\nu}^{\left(
2\right)  }\left(  z\right)  $ are the Hankel functions of the first kind and
the second kind, we have%
\begin{equation}
h_{l+\left(  n-3\right)  /2}^{\left(  1\right)  }\left(  kr\right)
=\frac{e^{ikr}}{kr}\left(  -i\right)  ^{l+\left(  n-1\right)  /2}\left(
-2ikr\right)  ^{l+\left(  n-1\right)  /2}U\left(  l+\frac{n-1}{2}%
,2l+n-1,-2ikr\right)  .
\end{equation}

Using large-distance asymptotics of the Tricomi confluent hypergeometric
function, $U\left(  a,b,z\right)  \overset{r\rightarrow\infty}{\sim}z^{-a}$,
we arrive at%

\begin{equation}
h_{l+\left(  n-3\right)  /2}^{\left(  1\right)  }\left(  kr\right)
\overset{r\rightarrow\infty}{\sim}\frac{e^{ikr}}{kr}\left(  -i\right)
^{l+\left(  n-1\right)  /2}. \label{hjianjin}%
\end{equation}
The $n$-dimensional scattering condition (\ref{biantiaojian}) becomes%
\begin{equation}
\psi\left(  r,\theta\right)  \overset{r\rightarrow\infty}{\sim}e^{ikr\cos
\theta}+f\left(  \theta\right)  \frac{e^{ikr}}{r^{\left(  n-1\right)  /2}},
\end{equation}
where the $n$-dimensional scattering amplitude
\begin{equation}
f\left(  \theta\right)  =\sum_{l=0}^{\infty}a_{l}\left(  \theta\right)
\frac{\left(  -i\right)  ^{l+\left(  n-1\right)  /2}}{k}.
\end{equation}
By eq. (\ref{al}), the large-distance asymptotic $n$-dimensional scattering
amplitude can be expressed as%
\begin{equation}
f\left(  \theta\right)  =\frac{1}{2ik}\frac{\left(  -i\right)  ^{\left(
n-3\right)  /2}\Gamma\left(  n/2-1\right)  }{\sqrt{\pi}\left(  k/2\right)
^{\left(  n-3\right)  /2}}\sum_{l=0}^{\infty}\left(  2l+n-2\right)  \left(
e^{2i\delta_{l}}-1\right)  C_{l}^{n/2-1}\left(  \cos\theta\right)  .
\end{equation}
The large-distance asymptotic $n$-dimensional differential scattering cross
section, by eqs. (\ref{hjianjin}) and (\ref{sigma}), reads%
\begin{align}
&  \frac{d\sigma}{d\Omega}\overset{r\rightarrow\infty}{\sim}r^{n-1}\frac
{1}{2ik}\sum_{l=0}^{\infty}\sum_{l^{\prime}=0}^{\infty}a_{l}^{\ast}\left(
\theta\right)  a_{l^{\prime}}\left(  \theta\right)  W_{r}\left[  \frac
{1}{r^{\left(  n-3\right)  /2}}\frac{e^{-ikr}}{kr}i^{l+\left(  n-1\right)
/2},\frac{1}{r^{\left(  n-3\right)  /2}}\frac{e^{ikr}}{kr}\left(  -i\right)
^{l^{\prime}+\left(  n-1\right)  /2}\right] \nonumber\\
&  =\sum_{l=0}^{\infty}\sum_{l^{\prime}=0}^{\infty}\frac{i^{l+\left(
n-1\right)  /2}a_{l}^{\ast}\left(  \theta\right)  }{k}\frac{\left(  -i\right)
^{l^{\prime}+\left(  n-1\right)  /2}a_{l^{\prime}}\left(  \theta\right)  }%
{k}\nonumber\\
&  =\left\vert f\left(  \theta\right)  \right\vert ^{2}.
\end{align}
Integrating the differential scattering cross section gives the large-distance
asymptotic $n$-dimensional total cross section:%
\begin{equation}
\sigma=\frac{4\pi^{\left(  n-1\right)  /2}}{\Gamma\left(  \left(  n-1\right)
/2\right)  }\frac{1}{k^{n-1}}\sum_{l=0}^{\infty}\left(  2l+n-2\right)  \left(
l+1\right)  _{n-3}\sin^{2}\delta_{l}.
\end{equation}

\section{Conclusions \label{Conclusions}}

In this paper, without large-distance asymptotics, we establish a
$n$-dimensional scattering theory.

According to Euler's scheme, functions can be classified by their asymptotics
\cite{blanton2012introduction}. Therefore, for scatterings of short-range
potentials, the scattering wave function is classified by the solution of the
asymptotic equation of the Schr\"{o}dinger equation, i.e., the Schr\"{o}dinger
equation with $V\left(  r\right)  =0$. That is to say, the solution of the
asymptotic equation is the asymptotics of the scattering wave function. In
conventional scattering theory, however, the solution of the asymptotic
equation is replaced by an approximate asymptotic solution of the asymptotic
equation. What we do in the preceding work \cite{liu2014scattering} (three
dimensions) and in the present paper (arbitrary dimensions) is to provide an
accurate scattering theory in which the scattering wave function is restricted
by the exact solution of the asymptotic equation rather than by an approximate
solution of the asymptotic equation.

An arbitrary-dimensional scattering theory has special importance in
renormalization of the scattering spectral method
\cite{graham2009spectral,pang2012relation}. In further research, we will
discuss the application of the arbitrary-dimensional
without-large-distance-asymptotics scattering theory to renormalization in the
scattering spectral method.

In particular, it is often useful to consider a physical problem in arbitrary
dimensions. Arbitrary-dimensional scattering problems have been considered in
many aspects: scatterings of the Fermi pseudopotential
\cite{wodkiewicz1991fermi}, scatterings of long-range potentials
\cite{bolle1984scattering}, and scatterings of black holes
\cite{cardoso2006black}. Besides scattering problems, there are also many
arbitrary-dimensional theories. In quantum mechanics and statistic physics,
one considers the $n+4$-dimensional scalar spheroidal harmonics
\cite{berti2006eigenvalues}, the\ Landau problem in even-dimensional space
$CP^{k}$ \cite{karabali2002quantum}, and Fermi gases in arbitrary dimensions
\cite{valiente2012universal}; in field theory, one considers Yang-Mills and
gravity theories in arbitrary dimensions \cite{cachazo2014scattering}; in
gravity and cosmology theories, one considers the thermodynamic curvature of
the Kerr and Reissner-Nordstr\"{o}m (RN) black holes in arbitrary dimensions
\cite{aaman2006geometry}, an arbitrary-dimensional gravitational theory with a
negative cosmological constant, arbitrary-dimensional \textit{AdS} Black Holes
\cite{belhaj2015heat}, and arbitrary-dimensional \textit{AdS} black branes
\cite{acena2014hairy}.

Starting from the result given by the present paper and the preceding work
\ \cite{liu2014scattering}, one can reconsider many scattering related
problems. The analytic property of scattering amplitudes, which used to be
treated by conventional scattering theory
\cite{arnecke2008jost,willner2006low,laha2005off,decanini2003complex,costa2012conformal}%
, can be now discussed without large-distance asymptotics in arbitrary
dimensions. An arbitrary-dimensional Lippmann-Schwinger equation without
large-distance asymptotics can be constructed in the frame of the scattering
theory given in the present paper. An arbitrary-dimensional vector and tensor
scattering theories without large-distance asymptotics can be established,
e.g., electromagnetic scatterings and gravitational wave scatterings, which
are usually studied in the frame of conventional scattering theory
\cite{pike2001scattering,colton2012inverse,dolan2008scattering,giddings2010gravitational}%
. The acoustic scattering is a scalar scattering. Two- and three- dimensional
acoustic scattering theories without large-distance asymptotics can also be
established. In conventional acoustic scattering theory, large-distance
asymptotics is imposed \cite{berthet2003using,ikehata2012inverse,liu2013near}
and, therefore, the information of the distance between target and observer is
lost. A very important application is to consider inverse scattering problems
without large-distance asymptotics in arbitrary dimensions; this is a
fundamental problem and is studied under large-distance asymptotics
\cite{chadan1997introduction,sabatier2000past}. Based on the relation of two
important quantum field theory methods \cite{dai2009number,pang2012relation},
the scattering spectrum method
\cite{rahi2009scattering,forrow2012variable,bimonte2012exact} and the
heat-kernel method
\cite{vassilevich2003heat,fucci2012heat,moral2012derivative,dai2010approach},
we establish a heat-kernel method for calculating phase shifts
\cite{li2015heat}. This result can be generalized to arbitrary-dimensional
cases. The result given by\ ref. \cite{liu2014scattering} and\ the present
paper can be also applied to scatterings of long-range potentials, which is
usually studied under large-distance asymptotics \cite{hod2013scattering}. In
particular, we will discuss the scattering of a wave on a black hole in
arbitrary dimensions, while research in literature is usually under
large-distance asymptotics
\cite{crispino2009electromagnetic,doran2002perturbation,dolan2006fermion,crispino2009scattering,raffaelli2013scattering}%
. String theory related scatterings \cite{chen2007quantum} can be also
considered without large-distance asymptotics.

%\appendix
%\section{Some title}
%Please always give a title also for appendices.

\acknowledgments

We are very indebted to Dr G. Zeitrauman for his encouragement. This work is supported in part by NSF of China under Grant
No. 11575125 and No. 11375128.

%\acknowledgments%ÖÂл
%%%%%%%%%%ÕýÎĽáÊø

%\begin{thebibliography}{99}

%\end{thebibliography}\endgroup

%\bibitem{a}
%Author, \emph{Title}, \emph{J. Abbrev.} {\bf vol} (year) pg.

%\bibitem{b}
%Author, \emph{Title},
%arxiv:1234.5678.

%\bibitem{c}
%Author, \emph{Title},
%Publisher (year).

% Please avoid comments such as "For a review'', "For some examples",
% "and references therein" or move them in the text. In general,
% please leave only references in the bibliography and move all
% accessory text in footnotes.

% Also, please have only one work for each \bibitem.

%\end{thebibliography}

\providecommand{\href}[2]{#2}\begingroup\raggedright\endgroup %(ÒýÓÃbblʱÓÃÕâ¾ä)

%\bibliographystyle{JHEP}
%\bibliography{refs}% Produces the bibliography via BibTeX.

\begin{thebibliography}{10} %(ÒýÓÃbblʱÓÃÕâ¾ä)
%bblµÄ´¿ÎÄÏײ¿·Ö·ÅÔÚÕâ
\bibitem{liu2014scattering}
T.~Liu, W.-D. Li, and W.-S. Dai, {\it Scattering theory without large-distance
  asymptotics},  {\em Journal of High Energy Physics} {\bf 2014} (2014), no.~6
  1--12.

\bibitem{ballentine1998quantum}
L.~E. Ballentine, {\em Quantum Mechanics: a modern development}.
\newblock World Scientific Publishing Company, 1998.

\bibitem{graham2009spectral}
N.~Graham, M.~Quandt, and H.~Weigel, {\em Spectral methods in quantum field
  theory}, vol.~777.
\newblock Springer Science \& Business Media, 2009.

\bibitem{pang2012relation}
H.~Pang, W.-S. Dai, and M.~Xie, {\it Relation between heat kernel method and
  scattering spectral method},  {\em The European Physical Journal C} {\bf 72}
  (2012), no.~5 1--13.

\bibitem{li2015heat}
W.-D. Li and W.-S. Dai, {\it Heat-kernel approach for scattering},  {\em The
  European Physical Journal C} {\bf 75} (2015), no.~6.

\bibitem{olver2010nist}
F.~W. Olver, D.~W. Lozier, R.~F. Boisvert, and C.~W. Clark, {\em NIST handbook
  of mathematical functions}.
\newblock Cambridge University Press, 2010.

\bibitem{joachain1975quantum}
C.~J. Joachain, {\em Quantum collision theory}.
\newblock North-Holland Publishing Company, Amsterdam, 1975.

\bibitem{blanton2012introduction}
L.~Euler and T.~by~John D.~Blanton, {\em Introduction to Analysis of the
  Infinite}.
\newblock No.~II. Springer, New York, 1990.

\bibitem{wodkiewicz1991fermi}
K.~W{\'o}dkiewicz, {\it Fermi pseudopotential in arbitrary dimensions},  {\em
  Physical Review A} {\bf 43} (1991), no.~1 68.

\bibitem{bolle1984scattering}
D.~Boll{\'e} and F.~Gesztesy, {\it Scattering observables in arbitrary
  dimension n¡Ý 2},  {\em Physical Review A} {\bf 30} (1984), no.~3 1279.

\bibitem{cardoso2006black}
V.~Cardoso, M.~Cavaglia, and L.~Gualtieri, {\it Black hole particle emission in
  higher-dimensional spacetimes},  {\em Physical review letters} {\bf 96}
  (2006), no.~7 071301.

\bibitem{berti2006eigenvalues}
E.~Berti, V.~Cardoso, and M.~Casals, {\it Eigenvalues and eigenfunctions of
  spin-weighted spheroidal harmonics in four and higher dimensions},  {\em
  Physical Review D} {\bf 73} (2006), no.~2 024013.

\bibitem{karabali2002quantum}
D.~Karabali and V.~P. Nair, {\it Quantum hall effect in higher dimensions},
  {\em Nuclear Physics B} {\bf 641} (2002), no.~3 533--546.

\bibitem{valiente2012universal}
M.~Valiente, N.~T. Zinner, and K.~M{\o}lmer, {\it Universal properties of fermi
  gases in arbitrary dimensions},  {\em Physical Review A} {\bf 86} (2012),
  no.~4 043616.

\bibitem{cachazo2014scattering}
F.~Cachazo, S.~He, and E.~Y. Yuan, {\it Scattering of massless particles in
  arbitrary dimensions},  {\em Physical review letters} {\bf 113} (2014),
  no.~17 171601.

\bibitem{aaman2006geometry}
J.~E. {\AA}man and N.~Pidokrajt, {\it Geometry of higher-dimensional black hole
  thermodynamics},  {\em Physical Review D} {\bf 73} (2006), no.~2 024017.

\bibitem{belhaj2015heat}
A.~Belhaj, M.~Chabab, H.~El~Moumni, K.~Masmar, M.~Sedra, and A.~Segui, {\it On
  heat properties of ads black holes in higher dimensions},  {\em Journal of
  High Energy Physics} {\bf 2015} (2015), no.~5 1--13.

\bibitem{acena2014hairy}
A.~Ace{\~n}a, A.~Anabal{\'o}n, D.~Astefanesei, and R.~Mann, {\it Hairy planar
  black holes in higher dimensions},  {\em Journal of High Energy Physics} {\bf
  2014} (2014), no.~1 1--21.

\bibitem{arnecke2008jost}
F.~Arnecke, J.~Madro{\~n}ero, and H.~Friedrich, {\it Jost functions and
  singular attractive potentials},  {\em Physical Review A} {\bf 77} (2008),
  no.~2 022711.

\bibitem{willner2006low}
K.~Willner and F.~A. Gianturco, {\it Low-energy expansion of the jost function
  for long-range potentials},  {\em Physical Review A} {\bf 74} (2006), no.~5
  052715.

\bibitem{laha2005off}
U.~Laha and B.~Kundu, {\it Off-shell jost solution for scattering by a coulomb
  field},  {\em Physical Review A} {\bf 71} (2005), no.~3 032721.

\bibitem{decanini2003complex}
Y.~D{\'e}canini, A.~Folacci, and B.~Jensen, {\it Complex angular momentum in
  black hole physics and quasinormal modes},  {\em Physical Review D} {\bf 67}
  (2003), no.~12 124017.

\bibitem{costa2012conformal}
M.~S. Costa, V.~Goncalves, and J.~Penedones, {\it Conformal regge theory},
  {\em Journal of High Energy Physics} {\bf 2012} (2012), no.~12 1--50.

\bibitem{pike2001scattering}
E.~R. Pike and P.~C. Sabatier, {\em Scattering, Two-Volume Set: Scattering and
  inverse scattering in Pure and Applied Science}.
\newblock Academic press, 2001.

\bibitem{colton2012inverse}
D.~Colton and R.~Kress, {\em Inverse Acoustic and Electromagnetic Scattering
  Theory}.
\newblock Applied Mathematical Sciences. Springer New York, 2012.

\bibitem{dolan2008scattering}
S.~R. Dolan, {\it Scattering and absorption of gravitational plane waves by
  rotating black holes},  {\em Classical and Quantum Gravity} {\bf 25} (2008),
  no.~23 235002.

\bibitem{giddings2010gravitational}
S.~B. Giddings and R.~A. Porto, {\it The gravitational s matrix},  {\em
  Physical Review D} {\bf 81} (2010), no.~2 025002.

\bibitem{berthet2003using}
R.~Berthet and C.~Coste, {\it Using a partial-wave method for sound--mean-flow
  scattering problems},  {\em Physical Review E} {\bf 67} (2003), no.~3 036604.

\bibitem{ikehata2012inverse}
M.~Ikehata, {\it An inverse acoustic scattering problem inside a cavity with
  dynamical back-scattering data},  {\em Inverse Problems} {\bf 28} (2012),
  no.~9 095016.

\bibitem{liu2013near}
H.~Liu, {\it On near-cloak in acoustic scattering},  {\em Journal of
  Differential Equations} {\bf 254} (2013), no.~3 1230--1246.

\bibitem{chadan1997introduction}
K.~Chadan, D.~Colton, L.~P{\"a}iv{\"a}rinta, and W.~Rundell, {\em An
  introduction to inverse scattering and inverse spectral problems}, vol.~2.
\newblock Siam, 1997.

\bibitem{sabatier2000past}
P.~C. Sabatier, {\it Past and future of inverse problems},  {\em Journal of
  Mathematical Physics} {\bf 41} (2000), no.~6 4082--4124.

\bibitem{dai2009number}
W.-S. Dai and M.~Xie, {\it The number of eigenstates: counting function and
  heat kernel},  {\em Journal of High Energy Physics} {\bf 2009} (2009), no.~02
  033.

\bibitem{rahi2009scattering}
S.~J. Rahi, T.~Emig, N.~Graham, R.~L. Jaffe, and M.~Kardar, {\it Scattering
  theory approach to electrodynamic casimir forces},  {\em Physical Review D}
  {\bf 80} (2009), no.~8 085021.

\bibitem{forrow2012variable}
A.~Forrow and N.~Graham, {\it Variable-phase s-matrix calculations for
  asymmetric potentials and dielectrics},  {\em Physical Review A} {\bf 86}
  (2012), no.~6 062715.

\bibitem{bimonte2012exact}
G.~Bimonte and T.~Emig, {\it Exact results for classical casimir interactions:
  Dirichlet and drude model in the sphere-sphere and sphere-plane geometry},
  {\em Physical review letters} {\bf 109} (2012), no.~16 160403.

\bibitem{vassilevich2003heat}
D.~V. Vassilevich, {\it Heat kernel expansion: user's manual},  {\em Physics
  Reports} {\bf 388} (2003), no.~5 279--360.

\bibitem{fucci2012heat}
G.~Fucci and K.~Kirsten, {\it Heat kernel coefficients for laplace operators on
  the spherical suspension},  {\em Communications in Mathematical Physics} {\bf
  314} (2012), no.~2 483--507.

\bibitem{moral2012derivative}
F.~Moral-G{\'a}mez and L.~Salcedo, {\it Derivative expansion of the heat kernel
  at finite temperature},  {\em Physical Review D} {\bf 85} (2012), no.~4
  045019.

\bibitem{dai2010approach}
W.-S. Dai and M.~Xie, {\it An approach for the calculation of one-loop
  effective actions, vacuum energies, and spectral counting functions},  {\em
  Journal of High Energy Physics} {\bf 2010} (2010), no.~6 1--29.

\bibitem{hod2013scattering}
S.~Hod, {\it Scattering by a long-range potential},  {\em Journal of High
  Energy Physics} {\bf 2013} (2013), no.~9 1--11.

\bibitem{crispino2009electromagnetic}
L.~C. Crispino, S.~R. Dolan, and E.~S. Oliveira, {\it Electromagnetic wave
  scattering by schwarzschild black holes},  {\em Physical review letters} {\bf
  102} (2009), no.~23 231103.

\bibitem{doran2002perturbation}
C.~Doran and A.~Lasenby, {\it Perturbation theory calculation of the black hole
  elastic scattering cross section},  {\em Physical Review D} {\bf 66} (2002),
  no.~2 024006.

\bibitem{dolan2006fermion}
S.~Dolan, C.~Doran, and A.~Lasenby, {\it Fermion scattering by a schwarzschild
  black hole},  {\em Physical Review D} {\bf 74} (2006), no.~6 064005.

\bibitem{crispino2009scattering}
L.~C. Crispino, S.~R. Dolan, and E.~S. Oliveira, {\it Scattering of massless
  scalar waves by reissner-nordstr{\"o}m black holes},  {\em Physical Review D}
  {\bf 79} (2009), no.~6 064022.

\bibitem{raffaelli2013scattering}
B.~Raffaelli, {\it A scattering approach to some aspects of the schwarzschild
  black hole},  {\em Journal of High Energy Physics} {\bf 2013} (2013), no.~1
  1--18.

\bibitem{chen2007quantum}
H.-Y. Chen, N.~Dorey, and R.~F.~L. Matos, {\it Quantum scattering of giant
  magnons},  {\em Journal of High Energy Physics} {\bf 2007} (2007), no.~09
  106.
\end{thebibliography}

\end{document}